\documentclass[aip,apl,reprint,showpacs,groupedaddress]{revtex4-1}
\usepackage{amssymb}
\usepackage{graphicx}
\usepackage{amsmath}

\usepackage{color}

\begin{document}

\setcounter{page}{1}

\title{Open-Circuit Voltage Deficit in Cu$_2$ZnSnS$_4$ Solar Cells by Interface Band Gap Narrowing}

\author{Ji-Sang Park}
\affiliation{Thomas Young Centre and Department of Materials, Imperial College London, Exhibition Road, London SW7 2AZ, UK}
\email[]{ji-sang.park@imperial.ac.uk}

\author{Sunghyun Kim}
\affiliation{Thomas Young Centre and Department of Materials, Imperial College London, Exhibition Road, London SW7 2AZ, UK}

\author{Samantha N. Hood}
\affiliation{Thomas Young Centre and Department of Materials, Imperial College London, Exhibition Road, London SW7 2AZ, UK}

\author{Aron Walsh}
\affiliation{Thomas Young Centre and Department of Materials, Imperial College London, Exhibition Road, London SW7 2AZ, UK 
                and Department of Materials Science and Engineering, Yonsei University, Seoul 03722, Korea}

\bibliographystyle{apsrev4-1}

\date{\today}
\begin{abstract}
There is evidence that interface recombination in Cu$_2$ZnSnS$_4$ solar cells contributes to the open-circuit voltage deficit. Our hybrid density functional theory calculations suggest that electron-hole recombination at the Cu$_2$ZnSnS$_4$/CdS interface is caused by a deeper conduction band that slows electron extraction. In contrast, the bandgap is not narrowed for the Cu$_2$ZnSnSe$_4$/CdS interface, consistent with a lower open-circuit voltage deficit.
\end{abstract}
\maketitle

Solar cells based on earth-abundant Cu$_2$ZnSn(S,Se)$_4$ (CZTSSe) absorber materials suffer from lower solar conversion efficiency than other mature technologies because of the large open-circuit voltage (V$_{\mathrm{OC}}$) deficit.\cite{polizzotti2013state,walsh2012kesterite,yang2016band,guchhait2016enhancement,bourdais2016cu,grenet2018analysis,yan2018cu}  
Since the open-circuit voltage is determined by the quasi-Fermi level splitting, understanding recombination mechanisms not only in the bulk region\cite{kim2018identification} but also at the interfaces\cite{Gunawan2010,AENM201301465,crovetto2017interface,redinger2018high} is essential to developing a proper passivation strategy to achieve a higher efficiency.\cite{repins2015effects}
In this circumstance, fundamental properties of interfaces formed in CZTSSe solar cells need to be thoroughly investigated,\cite{crovetto2017band,kaur2017strategic,antunez2017back,turnbull2018probing} a strategy that has proved effective in other mature technologies.\cite{Aberle2000,Battaglia2016,park2018point} 

One open question in this community is why there is stronger interface recombination in CZTSSe solar cells with a higher S composition ratio. This increased recombination is primarily characterized by a smaller activation energy for recombination compared with the bandgap energy.\cite{wang2010thermally,redinger2013influence,yan2018cu}
Previously, a cliff-type conduction band offset (CBO)\cite{crovetto2017band} has been suggested to be the culprit behind this stronger interface recombination.\cite{wang2010thermally} However, a recent study indicated that the CBO is instead actually a weak spike under the strain-free condition when temperature effects were considered.\cite{bartomeu2018}

Another recent model based on density functional theory (DFT) calculations insists that Cu-S bonds at the CZTS/CdS interface can introduce gap states 0.2$-$0.3 eV higher than the VBM even without point defects.\cite{crovetto2017interface}
Performing device simulations taking into account the interface states, the study was able to reproduce the experimentally measured temperature dependent V$_{\mathrm{OC}}$ data without cliff-type conduction band offsets.\cite{crovetto2017interface,crovetto2018large}
Since the electronic band gap of semiconductors is usually underestimated in the generalized gradient approximation (GGA) calculations,\cite{heyd2005energy} 
the Hubbard U correction, which has been widely used to improve the band gap and formation enthalpies,\cite{wang2006oxidation,jain2011formation,noh2011electronic} was applied in the study.\cite{crovetto2017interface}
The results are worth being re-examined by more advanced methods, considering the importance and implications of the conclusion.

In this letter, we report our first-principles DFT calculation results on the electronic structure of CZTS(Se)/CdS interface.
Several exchange-correlation functionals were examined to check how the interface band gap is narrowed.
We suggest that the interface band gap narrowing is caused by the lowered conduction band of CZTS. 
As the electron carriers are captured at the interface, the quasi-Fermi level splitting and the open-circuit voltage will be reduced significantly.

In our DFT calculations, we used the projector augmented wave (PAW) method to describe the interaction between ions and electrons,\cite{PAW} as implemented in the Vienna ab-initio Simulation Package (VASP) code.\cite{PhysRevB.54.11169}
To verify the effect of exchange-correlation functional on the electronic structure of materials, we employed various exchange-correlation functional including PBE,\cite{PBE} revised PBE,\cite{Zhang1998} PBEsol,\cite{Perdew2008} AM05,\cite{Armiento2005} SCAN,\cite{SCAN} and HSE06.\cite{HSE}
The wavefunctions were expanded in plane waves with an energy cutoff of 400 eV.
A 6$\times$6$\times$6 \textit{k}-point grid was used for Brillouin zone integration of the primitive cell.
Here we consider zinc-blende CdS (\textit{zb}-CdS), not thermodynamically stable wurtzite CdS (\textit{wz}-CdS) to avoid the effect of spontaneous polarization in the interface calculations.
Following our examination of the pristine interfaces, further work is needed to consider the role of defects and non-stoichiometry on the interfacial processes in kesterite solar cells.

\begin{figure}
\includegraphics[width=\columnwidth]{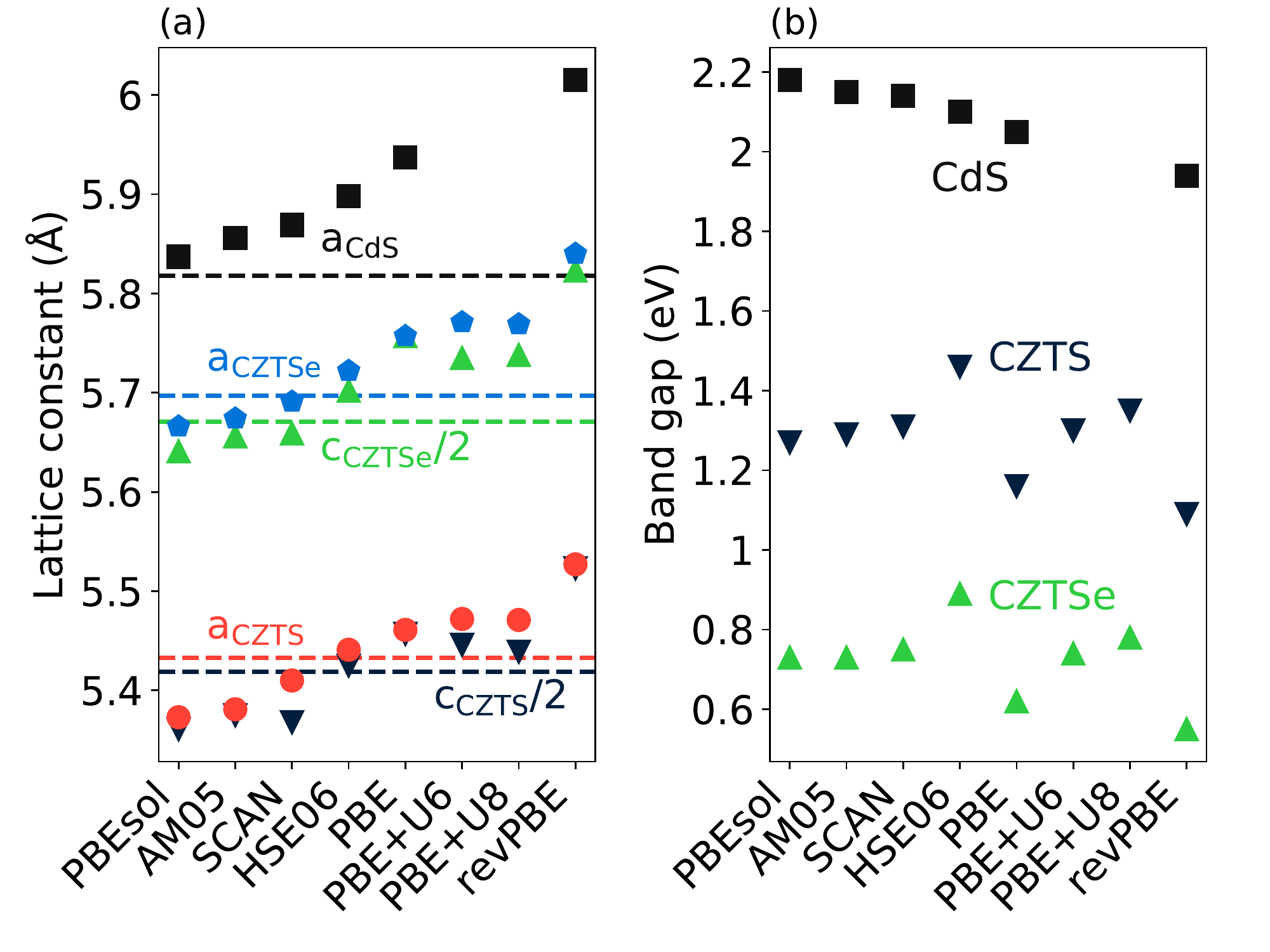}
\caption{\label{fig:1} (a) The lattice constants of zinc-blende CdS, kesterite CZTS, and kesterite CZTSe, optimized by using various exchange-correlation functionals. On-site coulomb potentials (6  or 8 eV) were applied to Cu $d$ and Zn $d$ orbitals. The experiment values\cite{Wei2000,Tobbens2016,bosson2016crystal} were denoted by horizontal dashed lines. (b) The band gap of each structure optimized by each exchange-correlation functional was re-calculated using the HSE06 functional.}
\end{figure}

The optimized lattice constants are summarized in Figure 1a.
Since CZTS and CZTSe have tetragonal symmetry, there are two lattice constants, which are the lattice constant along (100) and (010) directions, $a$, and that along (001) direction, $c$.
The experimentally measured lattice constants (dashed horizontal lines in Figure 1a)\cite{Tobbens2016,bosson2016crystal} are similar to the lattice constants optimized by the SCAN and the HSE06 functionals.
The SCAN functional seems better than the PBE functional for the investigation of the CZTS(Se)/CdS interfaces as the PBE overestimates the lattice constant of \textit{zb}-CdS more than the SCAN does.
We also note that $c/2$ is calculated to be generally lower than $a$, consistent with the experiment results.\cite{Tobbens2016,bosson2016crystal}
It is also noteworthy that the exchange-correlation functionals are arranged equally in all materials when the optimized lattice constants are sorted with increasing order.

Heavy computational cost of the hybrid calculation can be relieved if the internal coordinates are optimized at the GGA level only.\cite{park2018mechanism}
To find which functional is the most suitable method for this strategy, we obtained the band gap of CdS, CZTS, and CZTSe using the HSE06 functional keeping the structures optimized by various exchange-correlation functionals, as summarized in Figure 1b.
The SCAN seems superior to other functionals as it reproduces the closer band gap to the HSE06 value.  
The band gap can be improved by applying moderate on-site Coulomb potentials, however, larger deviations in the lattice constants were obtained.
The result indicates that we can describe the electronic structure cost-effectively by relaxation of the atomic structure using SCAN functional and a subsequent SCF calculation using the HSE06 to correct the band gap of CZTS(Se).

\begin{figure}
\includegraphics[width=\columnwidth]{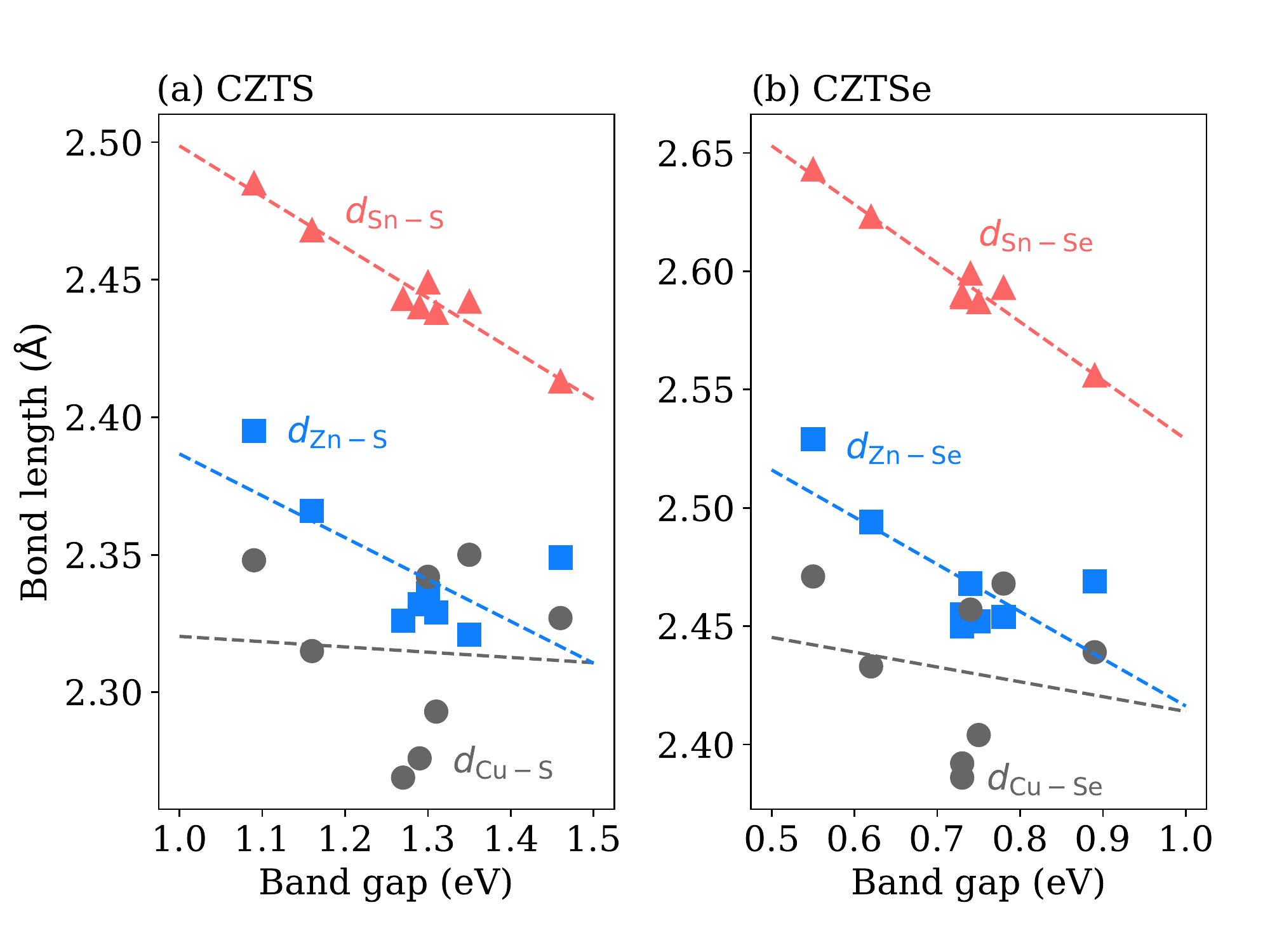}
\caption{\label{fig:2} Variation in the calculated HSE06 bandgap of (a) CZTS and (b) CZTSe with respect to the bond lengths in the underlying crystal structure models.  }
\end{figure}

There is no such clear correlation between the lattice constants and the HSE06 band gap of CZTS(Se) summarized in Figure 1b. 
But the change of the band gap is well explained by the change of the Sn-S(Se) bond length (Figure 2), indicating that the internal coordinates play a significant role in determining the band gap.
The conduction band minimum (CBM) of CZTS(Se) is an anti-bonding state of Sn \textit{s} and S(Se) \textit{p} orbitals, thus the shortened Sn-S(Se) bond length results in the larger band gap.\cite{park2015ordering}
The average Zn-S(Se) bond length also follows the same trend with larger deviations, whereas the Zn atoms do not constitute the band edges.
The low band gap calculated by using the PBE optimized structure is explained by the lengthened Sn-S(Se) bond by 0.05 {\AA} (0.07 {\AA}).
Therefore, if the experiment band gap is reproduced by applying large on-site Coulomb potential on Cu \textit{d} orbitals, then the valence band maximum, which is an anti-bonding state of Cu \textit{d} and S(Se) \textit{p} orbitals, could be too deep with respect to the vacuum level within PBE+U calculations.
The band gap of \textit{zb}-CdS increases monotonically with decreasing of the lattice constant because the Cd-S bond length also changes accordingly.

The atomic structure of a (100) CZTS/(100) \textit{zb}-CdS interface model is shown in Figure 3a.
The (100) Miller index of CdS is justified by experimental evidence of the epitaxial growth of CdS on CZTS.\cite{tajima2014atom,liu2016nanoscale}
The supercell of the interface model includes six CZTS double layers and the same number of \textit{zb}-CdS double layers.
The two lattice vectors parallel to the interface plane were set to [0,$a$,0] and [0,0,$c$], where the lattice constants $a$ and $c$ are those obtained using the HSE06.
A perfectly clean CZTS/CdS interface is expected to have states composed of Cu-S, Zn-S, Sn-S, and/or Cd-S states, in principle.
Therefore, the position of the interface states should be affected by the distance between the CZTS layer and the adjacent CdS.
To avoid artificially close or distant layers, the cell size along the direction normal to the interface was optimized within 0.01 {\AA}.
In each calculation, the CZTS layers were fixed because otherwise it results in the lower HSE06 band gap. 
The other layers (CdS layers and S atoms at the interfaces) were relaxed using the PBEsol functional until the residual force becomes smaller than 0.03 eV {\AA}$^{-1}$ because the functional reproduces the atomic structure of \textit{zb}-CdS most (see Figure 1a).
A (100) CZTSe/(100) \textit{zb}-CdS interface model was generated similarly.

\begin{figure}
\includegraphics[width=\columnwidth]{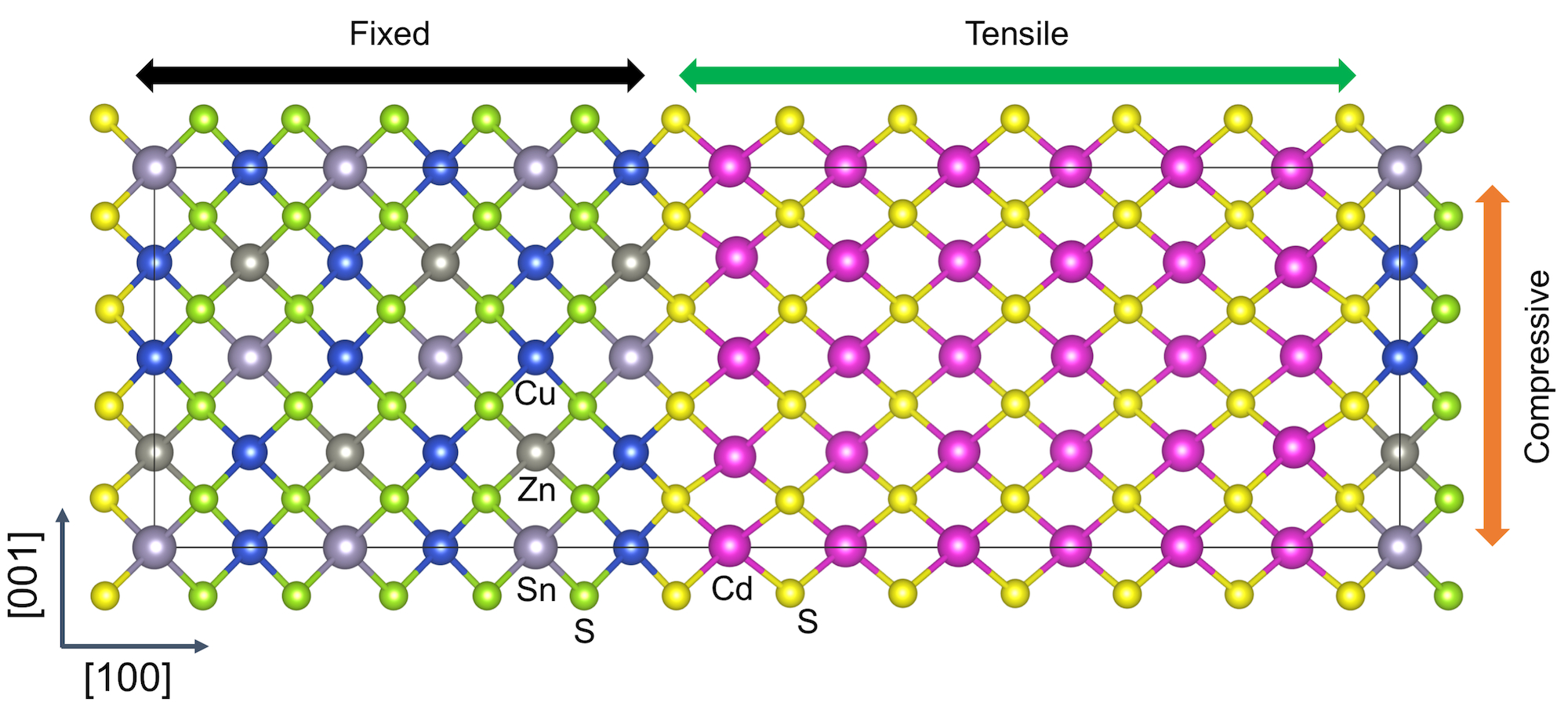}
\caption{\label{fig:3} Atomic structure of a (100) CZTS/(100) CdS interface model. S atoms with green color were fixed in position, and the other S atoms were freely relaxed when the cell size along the interface normal direction was optimized.}
\end{figure}

Since CdS has the larger lattice constant than CZTS, a biaxial compressive strain is applied to the CdS layers, and thus the lattice constant along the [100] direction is elongated. 
When the CdS atoms are optimized with the PBEsol functional, the lattice constant along the [100] direction is increased by 8.3 \% in the CZTS/CdS interface model compared to the HSE06 optimized lattice constant (5.90 {\AA}).
Much smaller change (1.7 \%) is observed in the CZTSe/CdS interface model because of the smaller difference in the lattice constant.
This large lattice mismatch between CZTS and CdS will make epitaxial growth difficult in large areas. The epitaxial growth observed in the experiment is probably due to the nanocrystalline nature of CdS.\cite{wuu1990indium,liu2016nanoscale}

After we determined the size of the supercell, we optimized the internal coordinates including the CZTS(Se) layers using PBE+U or SCAN+U functionals.
Since it is computationally heavy to relax the interface structures using the HSE06 functional, we performed a single SCF calculation with the HSE06 functional using the optimized structures by PBE+U or SCAN+U functionals.
The band edges of bulk CZTS(Se) were estimated using the local potential as a reference.
The highest occupied states and the lowest unoccupied states in the interface calculations were compared to the estimated band edges.

\begin{figure}
\includegraphics[width=\columnwidth]{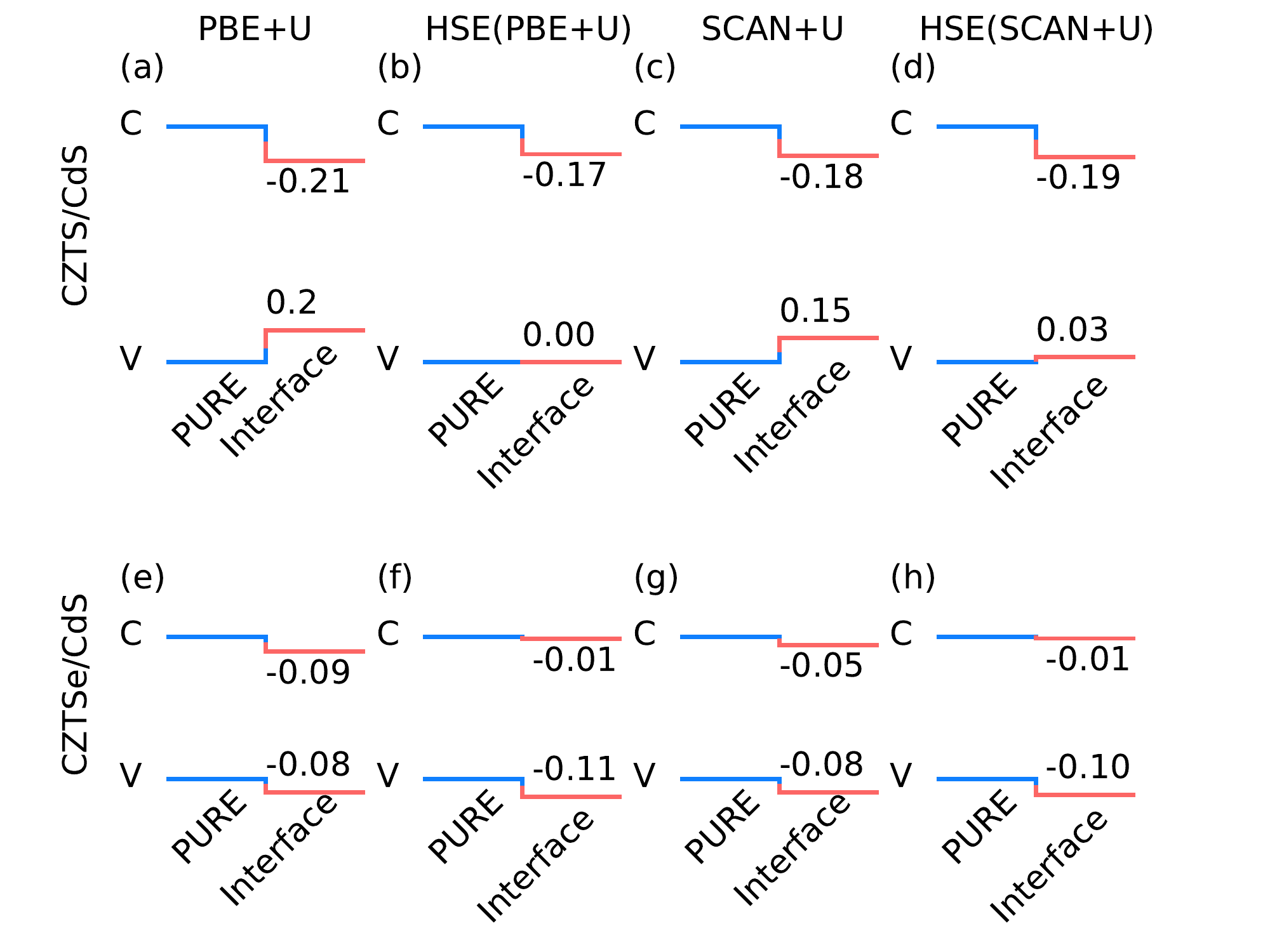}
\caption{\label{fig:4} The valence (V) and conduction (C) band edges of bulk CZTS(Se) and those from the interface calculations. We performed HSE06 calculations to obtain the band offset diagrams in the second and the fourth columns while the structure optimized by PBE+U and SCAN+U were used, respectively. The offset values are given in eV.}
\end{figure}

Figure 4 shows the band edge positions calculated from the interface calculation (red) with respect to those obtained from the bulk calculations (blue).
Consistent with the previous result, we found the increase of the VBM at the CZTS/CdS interface when PBE+U (Figure 4a) or SCAN+U (Figure 4c) functionals were used.\cite{crovetto2017interface}
When the electronic structure was calculated using the HSE06 functional, however, there was no evidence of the increased VBM (Figure 4b and Figure 4d).
It is also worth emphasizing that the CBM is lowered in every calculation (0.17-0.21 eV) due to the lengthened Sn-S bonds at the interface. 
Such interface band gap narrowing is not clearly found in CZTSe/CdS interface as the CBM is composed of Sn-Se anti-bonding which is lower than Sn-S anti-bonding.

We also analyzed the electronic structure using the HSE06 functional after we optimized the CdS domain only.
In the CZTS/CdS calculation, the CBM of CZTS is lowered by 0.18 eV while the VBM is increased by only 0.01 eV.
On the other hand in the CZTSe/CdS calculation, the CBM and the VBM are decreased by 0.03 eV and 0.09 eV, respectively. 
The CBM of CZTS is reduced no matter whether the CZTS layers were relaxed or not, and thus we rule out an argument that the interface band gap narrowing is caused by the structural relaxation.

We expect that our model is thick enough to confirm nature of the interface band gap narrowing as the electrostatic potential converges quickly as compared to the eigenvalues, which enables a cost-effective estimation of the band edges of pure CZTS(Se).\cite{park2018quick}
We also made another CZTS/CdS model by doubling the CZTS and CdS layers, performed a SCAN+U calculation, and found that the reduced CBM of 0.19 eV is reproduced.

We note that the VBM of CZTSe is reduced by $\sim$0.1 eV in every calculation as Cu-S bonds are formed at the interfaces in our model.
To quantitatively prove this explanation, we substituted Se atoms for S atoms at the interfaces and relaxed the substituted Se atoms using PBE+U functional.
The valence band offset (VBO) from the substituted interface was calculated to be -0.03 eV, which is clearly higher in absolute energy than the value before the substitution (-0.08 eV).
We note that a slightly higher S composition at the CZTSe/CdS interface could be beneficial as the hole barrier is formed at the interface.

\begin{figure}
\includegraphics[width=\columnwidth]{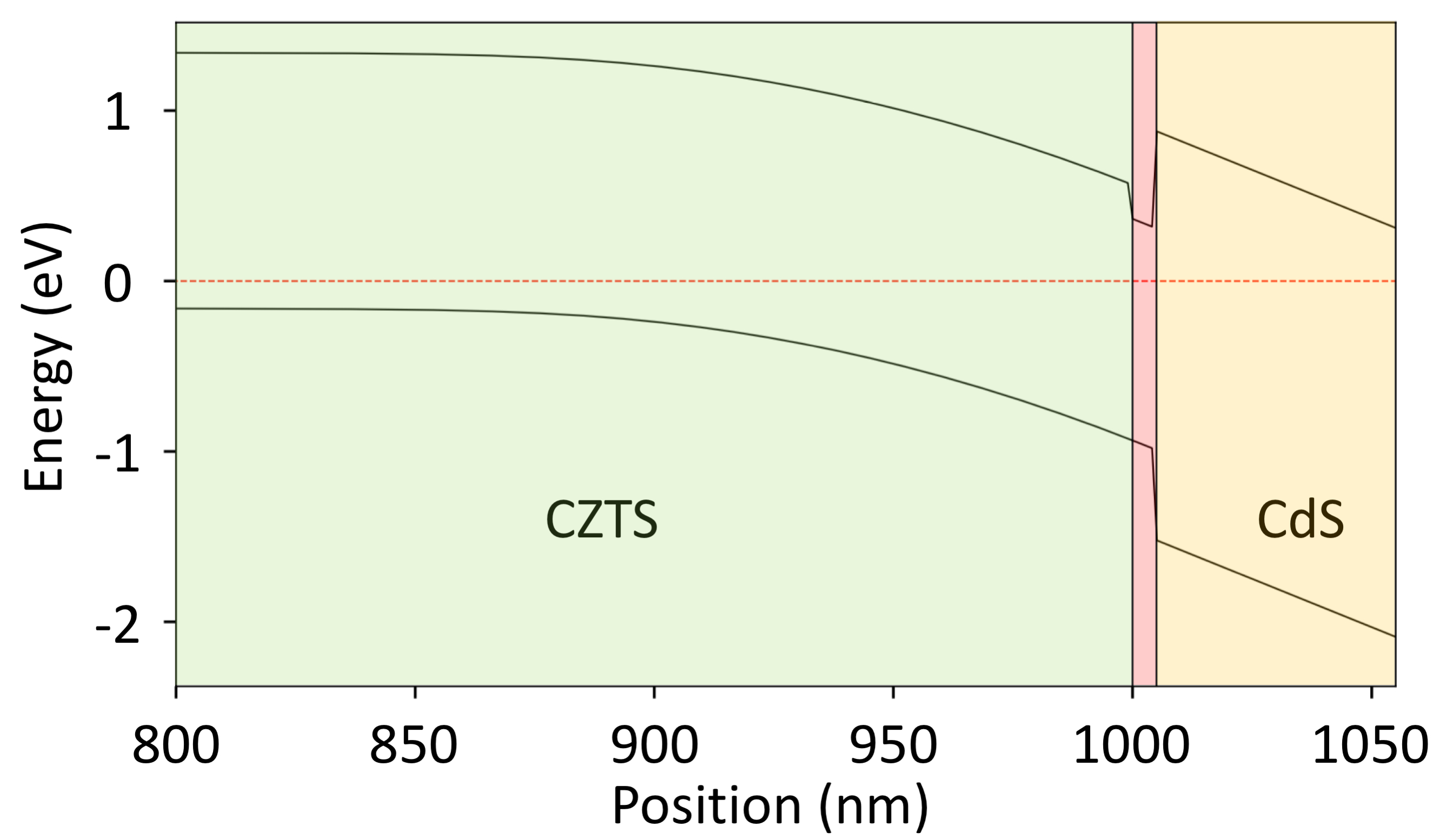}
\caption{\label{fig:5} The electronic band diagram of CZTS/CdS interface obtained using a Poisson-Boltzmann equation solver via finite differences. The red dotted lines represent the Fermi level, which is set to 0 eV. Position represents the distance from the back contact.}
\end{figure}

The band diagram of a CZTS/CdS interface was obtained by the solving the Poisson-Boltzmann equation as shown in Figure 5.
We adapted the parameters used in another study.\cite{crovetto2017interface}
Thin layers of CZTS ($\simeq$ 5 nm, red shaded region) is set to have lower conduction band than bulk CZTS by 0.2 eV because of the interface states.
The lowered CBM at the interface means that electron charge carriers could be accumulated at the CZTS/CdS interface.
The accumulated charges will recombine radiatively or through recombination centers, and then the quasi-Fermi level splitting, which corresponds to the open-circuit voltage, becomes narrower because of the resulting stronger interface recombination.\cite{kim2018identification,neuschitzer2018revealing}

Based on the calculation results, we suggest that the interface band gap narrowing by the lowered conduction band is an origin of the larger V$_{\mathrm{OC}}$ deficit in the CZTS solar cells.
Such interface band gap narrowing was not observed in CZTSe/CdS interface, and thus the V$_{\mathrm{OC}}$ is less affected by the interface recombination in CZTSe solar cells.
Consistent with this expectation, the activation energy for the dominant recombination measured from CZTSe solar cells is almost equivalent to the band gap of the absorber layer.\cite{redinger2013influence,neuschitzer2015optimization}
We also note that the open-circuit voltage deficit has been discussed comprehensively based on the band edge fluctuations.\cite{rau2004radiative,gokmen2013band,park2015ordering,park2018opposing} The band gap fluctuation model, the potential fluctuation model, and the current model based on the interface band gap narrowing do not contradict to each other because the fluctuation happens in bulk regions while the band gap narrowing occurs at the interface.

Experimental studies show that not only the bulk properties but also the interface recombination is essential to improve the CZTS solar cell.
The interface can be improved by inserting thin Al$_2$O$_3$ layers between CZTS(Se)/CdS interface,\cite{lee2016atomic} potentially due to less elemental intermixing.\cite{kim2017improving} 
CZTS solar cells with Zn$_{1-x}$Sn$_x$O$_y$ buffer layer also exhibit higher V$_{\mathrm{OC}}$.\cite{ericson2017zinc}
Ge doping was claimed to be effective to reduce the recombination, which mechanism should be investigated further.\cite{neuschitzer2018revealing}
Effect of the light soaking on the cell parameters could be investigated in future studies to widen our understanding.\cite{guo2010fabrication,neuschitzer2015optimization,chantana201820}

To investigate the electronic structure of CZTS/CdS interface, we employed (100) CZTS/(100) CdS interface. The biaxial tensile strain applied to the CdS layers does not introduce bending in the electrostatic potential, thus we used the potential far from the interface as a reference. When we generated an epitaxial (112) CZTS/(111) CdS interface model, the strong piezoelectric polarization was built along the interface normal direction, resulting in a linear slope in the electrostatic potential.

In summary, we re-examined the electronic structure of the CZTS(Se)/CdS interface and conclude that the bandgap narrowing is caused by a change in the conduction band energy rather than the valence band.
We find that a strong on-site Coulomb potential is required to reproduce the HSE06 band gap in GGA+U calculations, while the large on-site Coulomb potential can result in the error in the electron affinity.
We also obtained the lattice constants of kesterites and CdS using various exchange-correlation functionals, and suggest a way to reduce errors in the interface calculations. 

See supplementary material for the physical properties of CZTS and CZTSe calculated using various exchange-correlation functionals. The primary data for this article is available in a repository at https://zenodo.org/record/1478110. 

We thanks Andrea Crovetto and Mattias Lau N$\o$hr Palsgaard for helpful discussion.
This project has received funding from the European H2020 Framework Programme for research, technological development and demonstration under grant agreement no. 720907. See http://www.starcell.eu. 
Via our membership of the UK's HPC Materials Chemistry Consortium, which is funded by EPSRC (EP/L000202), this work used the ARCHER UK National Supercomputing Service (http://www.archer.ac.uk). We are grateful to the UK Materials and Molecular Modelling Hub for computational resources, which is partially funded by EPSRC (EP/P020194/1). 
JP thanks the Royal Society for a Shooter International Fellowship.

\bibliography{czts}

\end{document}